\documentclass{andp2012}
\usepackage[english]{babel}
\usepackage{color}
\pagespan{}{}
\category{Original Paper}
\subcategory{}
\title{{Exact Mobility edges and topological Anderson insulating phase in a slowly varying quasiperiodic model}}
\author[F. Author]{Zhanpeng Lu\inst{1}}
\author[F. Author]{Zhihao Xu\inst{1}}
\author[F. Author]{Yunbo Zhang\inst{2}
	\footnote{Corresponding author\quad E-mail:~\textsf{ybzhang@zstu.edu.cn}}}
\address[1]{Institute of Theoretical Physics and State Key Laboratory of Quantum Optics and Quantum Optics Devices, Shanxi University, Taiyuan 030006, China}
\address[2]{Key Laboratory of Optical Field Manipulation of Zhejiang Province and Physics Department of Zhejiang Sci-Tech University, Hangzhou 310018, China}
\begin{abstract}
We uncover the relationship of topology and disorder in a one-dimensional Su-Schrieffer-Heeger chain subjected to a slowly varying quasi-periodic modulation. By numerically calculating the disorder-averaged winding number and analytically studying the localization length of the zero modes, we obtain the topological phase diagram, which implies that the topological Anderson insulator (TAI) can be induced by a slowly varying quasi-periodic modulation. Moreover, unlike the localization properties in the TAI phase caused by random disorder, mobility edges can enter into the TAI region identified by the fractal dimension, the inverse participation ratio, and the spatial distributions of the wave functions, the boundaries of which coincide with our analytical results.
\end{abstract}
\shortabstract
\begin{document}
\maketitle
\section{\label{sec.1}Introduction}
Topological phases of matter have attracted broad interest over the past decades \cite{Thouless1982,Hasan2010,QXliangRMP2011,BansilRMP2016,Chiu2016,ArmitageRMP2018}. A topological insulator displays the gapped bulk states and the gapless edge modes lied in the bulk gap, which can be characterized by a non-trivially topological number. The Su-Schrieffer-Heeger (SSH) model is the simplest two-band topological system initially introduced to research the polyacetylene that exhibits rich physical phenomena \cite{Su1979}, such as fractional charge soliton excitation \cite{Jackiw1976,Heeger1988}and nontrivial edge modes \cite{Ganeshan2013}. Its chiral symmetry leads to nontrivial topology confirmed by a non-zero winding number and the emergence of the zero-energy edge modes under open boundary conditions (OBCs) \cite{Song2019PRL,XZhao2020PRA1,Bo2021}.

On the other hand, Anderson localization in a disordered medium  \cite{Anderson1958,Ramakrishnan1985RMP,Abanin2017,Huse2017} was first proposed in 1958, which has been realized in various experimental platforms such as cold atoms \cite{Billy2008Nat,Roati2008Nat} and microwaves cavity \cite{Chabanov2000Nat,Pradhan2000PRL}. Compared with traditional Anderson models, where even an infinitesimal random potential leads to localization in one- and two-dimensional cases, the quasi-periodic systems display distinctive localization properties. The one-dimensional Aubry-Andr\'e-Harper (AAH) model as one of the paradigmatic examples undergoes a metal-insulator transition characterized by the self-duality property when the strength of the quasi-periodic potential exceeds a finite critical value \cite{Aubry1980,Harper1955,Longhi2019PRL,Longhi2019PRB,Hongfu2021}. Remarkably, the presence of the mobility edge, one of the essential concepts in disordered systems seen only in the traditional three-dimensional Anderson model \cite{Mott1987JPC}, has been demonstrated in various generalized AAH models even in one dimensional systems. Some generalized AAH models displaying the mobility edges in compact analytic forms are proposed when the so-called self-dual symmetry is broken \cite{DasSarma2010PRL,Liu2020PRB,Wangg2020PRL,Tong2020PRB,Xu2021,Xu2021arxiv}, or the system is in the form of slowly varying modulations \cite{DasSarma1988PRL,DasSarma1990PRB,Tong2018PRB,Tong2018PLA}. {Recent advances in the mobility edges have been extended to the anomalous mobility edges separating the localized states from the critical ones in one-dimensional disordered systems \cite{TongLonghi2022}, which is different from the traditional mobility edges.}

Recently, great effort has been devoted to understand the interplay of topology and disorder, which brings new perspectives, such as the topological phase transition can be introduced by the quasi-periodic disorder in the one-dimensional SSH chain under the intercell hopping strength exceeds the intracell hopping strength \cite{Tong2018PLA}, and topological phase with critical localization consisting of only critically localized states \cite{Bo2021,Tong2021PRB,Huhui2016}.  One of the hallmark characteristics of a topological insulator is the robustness of the nontrivial edge states against weak disorder in the underlying lattice \cite{Hasan2010,Prodan2010}. When the amplitude of disorder is large enough, the topological features eventually disappear, accompanied by collapse of the nontrivial topological number \cite{Cai2013}. Conversely, a modest disorder added to a trivial system can lead to the emergence of protected edge modes and quantized topological charges. Such disorder-driven topological phase is named the topological Anderson insulator (TAI) \cite{Shen2009,Groth2009,Guo2010,Zhang2012,Song2012,Girschik2013,Hughes2014,Zhang2019,Hughes2021,zhanggq2021}. The experimental observation of the disorder-induced TAI has been reported in two-dimensional(2D) photonics \cite{Sttzer2018,Liu2020}, one-dimensional(1D) engineering synthetic 1D chiral symmetric wires with a precisely controllable random disorder \cite{Meier2018}. Recent advances of the TAI induced by a random disorder have been extended to non-Hermitian systems \cite{DZhang2020,Tangg2020} and spin-orbit coupled superconductors \cite{Borchmann2016,Hua2019}. On the other hand, the TAI phase is also shown to appear in the SSH model with a quasi-periodic modulated intercell hopping term with all the eigenstates being localized \cite{Longhi2020}.

Some interesting questions arise here: whether a slowly varying modulation can lead to a TAI phase, and what the localization properties in such case are. To address these, we consider a toy model, which is the SSH model subjected to a slowly varying intracell hopping and open boundary condition, focusing on the topology and localization features in the system. The topological phase diagrams will be obtained by numerically calculating the topological invariants in real space, i.e., the disorder-averaged winding number, and the topological phase transition points can be determined by the localization length of the zero modes. Moreover, the localization properties of the system will be explored by some proven means such as the inverse participation ratio, the fractal dimension, and the spatial distributions of the eigenstates. Specifically, we try to obtain analytic expressions for the mobility edges, if they exist.

The arrangement of the paper is as follows: In Sec.~\ref{sec.2}, we briefly introduce the Hamiltonian of the SSH model with a slowly varying intracell hopping term. In Sec.~\ref{sec.3}, we obtain the topological phase diagram and discuss the fate of topological zero-energy modes. Furthermore, we study the localization properties of the system in Sec.~\ref{sec.4}. Finally, a conclusion is presented in Sec.~\ref{sec.5}.
\section{\label{sec.2}Model and Hamiltonian}
\begin{figure}[tbp]
\includegraphics[width=0.45\textwidth]{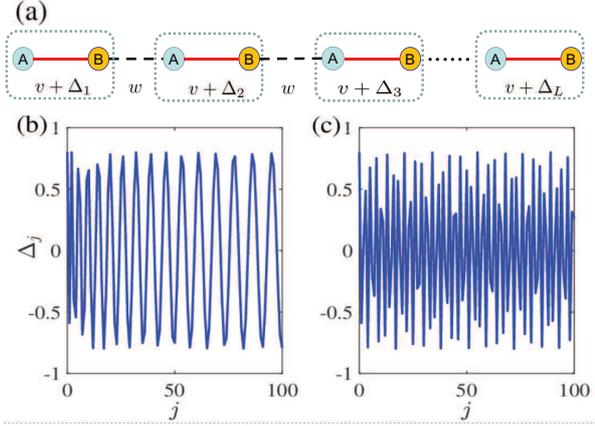}
\caption{(Color online) (a) Schematic of the SSH model of $L$ unit cells with a modulated intracell hopping term. The dotted box represents the unit cell.  $w$ and $v+\Delta_j$ represent the uniform intercell hopping and the $j$-dependent intracell hoppping, respectively. (b) Sketch of a slowly varying modulation $\Delta_j=\Delta \cos(2\pi\beta j^u+\phi)$ for $u=0.7$.
(c) Sketch of an AA-type modulation with $u=1$.
Here, $\beta=\frac{\sqrt{5}-1}{2}$, $\phi=0$, and $\Delta=0.8$.}
\label{Fig1}
\end{figure}

We consider a toy model with a slowly varying quasi-periodic modulation, which can be described by the Hamiltonian
\begin{eqnarray}
H&=&\sum_{j}v_j\left(c^{\dag}_{j,B}c_{j,A}+h.c\right)+w\sum_{j} \left(c^{\dag}_{j+1,A}c_{j,B}+h.c\right).
\label{eq1}
\end{eqnarray}
Here $c^{\dag}_{j,\alpha}$($c_{j,\alpha}$) creates (annihilates) a particle on the sublattice site
$A(B)$ in the $j$-th lattice cell, and $L$ is the length of the system.  As shown in \textbf{Figure~\ref{Fig1}}a, the intercell hopping strength $w$ is denoted by the black dashed line, which is set as the unit energy, \emph{i.e.}, $w=1$, and the $j$-dependent intracell hopping
\begin{eqnarray}
	v_j&=&v+\Delta_j,
	\label{eqa}
\end{eqnarray}
with
\begin{eqnarray}
	\Delta_j&=&{\Delta}\cos(2\pi\beta j^u+\phi),
	\label{eqb}
\end{eqnarray}
is denoted by the red solid line. Here $v$ is the site-independent hopping inside the unit cell, $\Delta$ is the strength of incommensurate modulation, $\beta$ is the modulation frequency, and $\phi\in[0,2\pi]$ is an arbitrary phase. In the thermodynamic limit, we have
\begin{eqnarray}
\lim_{j\to\infty}\left|\frac{d \Delta_j}{d j}\right|&=&\lim_{j\to\infty} 2\pi\Delta\beta u\frac{\left|\sin(2\pi\beta j^u+\phi)\right|}{j^{1-u}}\to 0,
\label{eq2}
\end{eqnarray}
with the slowly varying parameter $0<u<1$. The result indicates the difference of $\Delta_j$ tends to 0 when $j$ is large enough as shown in \textbf{Figure~\ref{Fig1}}b, which corresponds to a slowly varying modulation \cite{DasSarma1988PRL,DasSarma1990PRB}. For $\Delta=0$ or $u=0$, the system reduces to a standard SSH model \cite{Su1979} with a uniform intracell hopping amplitude. When the intercell hopping strength exceeds the intracell hopping strength, system undergoes a topological phase transition characterized by the emergence of  zero-energy edge modes and nontrivial winding number. For $\Delta\ne 0$ and $u=1$, $\Delta_j$ corresponds to an AA-type modulation illustrated in \textbf{Figure~\ref{Fig1}}c. The emergence of the TAI in SSH model with an AA-type modulation has been reported \cite{Longhi2020}. In the following, we testify the existence of the TAI in the SSH model with a slowly varying intracell modulation and discuss the corresponding localization properties. We fix $\beta=\frac{\sqrt{5}-1}{2}$ and $u=0.7$ for our discussion.

\section{\label{sec.3}Topological phase diagram}
A TAI is characterized by the emergence of the protected edge states and the quantized topological charge induced by the addition of sufficient disorder or incommensurate modulation to a trivial band structure. To detect the TAI phase, we first utilize the open-bulk winding number to characterize the topological properties for our slowly varying SSH model. For a given modulation configuration, we can diagonalize the open-chain Hamiltonian as $H|\psi^{n}\rangle=E_n|\psi^{n}\rangle$ and $H|\tilde{\psi}^{n}\rangle=-E_n|\tilde{\psi}^{n}\rangle$ to obtain a pair of chiral-symmetric partners $|\psi^{n}\rangle$ and $|\tilde{\psi}^{n}\rangle$ with the relation $|\tilde{\psi}^{n}\rangle =S|\psi^{n}\rangle$, where the entries of $S$ are $S_{i\alpha,j\gamma}=\delta_{ij}(\sigma_z)_{\alpha\gamma}$ with $i,j$ referring to the unit cell and $\alpha,\gamma$ to the sublattice. We introduce an open-boundary $Q$ matrix given by
\begin{equation}
Q=\sum_{n}\left(|\psi^{n}\rangle \langle \psi^{n}| - |\tilde{\psi}^{n}\rangle \langle \tilde{\psi}^{n}|\right),
\label{eq3}
\end{equation}
where $\sum_n$ is the sum over the eigenstates in the bulk spectrum without the edge modes. The open-bulk winding number in real space is defined as \cite{Song2019PRL}
\begin{equation}
W_c = \frac{1}{2L^{\prime}}\mathrm{Tr}^{\prime}(SQ[Q,X]).
\label{eq4}
\end{equation}
Here, $X$ is the coordinate operator, namely $X_{i\alpha,j\gamma}=j\delta_{ij}\delta_{\alpha\gamma}$. The length of the system $L$ can be divided into three intervals with length $l$, $L^{\prime}$ and $l$, \emph{i.e.}, $L=L^{\prime}+2l$. The symbol $\mathrm{Tr}^{\prime}$ represents the trace over the middle interval with length $L^{\prime}$. When $L \to \infty$, the open-bulk winding number method can well serve for the system deviated from periodicity, and $W_c$ is quantized to an integer, while a modest size is enough in the practical calculation. We also define the disorder-averaged winding number $\overline{W}=1/N_c\sum_{c=1}^{N_c} W_c$ with the configuration number $N_c$. Here, we take $N_c=100$ disorder realizations for different $\phi=[0,2\pi]$ throughout the work.

\begin{figure}[tbp]
\includegraphics[width=0.5\textwidth]{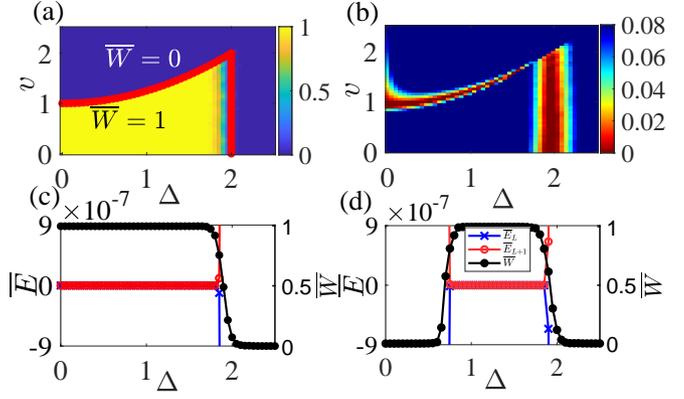}
\caption{(Color online) (a) Disorder-averaged winding number $\overline{W}$ as a function of the modulation strength $\Delta$ and the amplitude of the intracell hopping $v$ with $L=300$ and $N_c=100$ disorder realizations. The colorbar shows the value of disorder-averaged winding number $\overline{W}$. Two red dot lines represent the analytic critical lines Equation (\ref{eq9}) for the divergence of localization length $\lambda$. (b) $\lambda^{-1}$ as a function of $\Delta$ and $v$ for $L=600$ and $N_c=100$ disorder realizations. The colorbar shows the value of $\lambda^{-1}$. (c)(d) Two disorder-averaged energies $\overline{E}_L$ and $\overline{E}_{L+1}$ in the center of the spectrum, and the disorder-averaged winding number $\overline{W}$ as a function of the modulation strength $\Delta$ under OBCs with  $v=0.5$(c) and $v=1.1$(d), respectively.}
\label{Fig2}
\end{figure}

\textbf{Figure~\ref{Fig2}}a shows the topological phase diagram characterized by $\overline{W}$ versus the modulation strength $\Delta$ and the amplitude of the intracell hopping $v$. In the absence of the modulation $\Delta=0$, the standard SSH model exhibits a topological phase transition point at $v=1$ corresponding to a jump of $\overline{W}$ from $1$ to $0$ with the increase of $v$. When we turn on the slowly varying modulation $\Delta_j$ to the nontrivial band structure with $v<1$, the zero-energy modes are robust against the slowly varying modulation. With the increase of $\Delta \lesssim 2$, the two disorder-averaged zero modes $\overline{E}_L$ and $\overline{E}_{L+1}$ start to break into non-zero pairs, accompanied by the jump of the winding number to zero, as shown in \textbf{Figure~\ref{Fig2}}c for $v=0.5$ under OBCs. Here, $\overline{E}_{n}=1/N_c\sum_{c=1}^{N_c} E_n^c$ with $E_n^c$ being the $n$-th eigenenergy for a given modulation configuration. As shown in \textbf{Figure~\ref{Fig2}}a, we find that the topological transition point is independent of $v$ at $\Delta = 2$. It indicates that the topologically nontrivial phase will collapse as long as the strength of the slowly varying modulation reaches a critical value. When $v \in (1,2)$ the system remains a topologically trivial phase as in a standard SSH model for $\Delta>0$ and the TAI phase emerges in a finite regime between $v=\Delta^2/4+1$ and $\Delta=2$, which is characterized by the nontrivial winding number and the zero-energy edge modes. The deviation from the non-zero integer of the disorder-averaged winding numbers $\overline{W}$ near the phase transition boundaries comes from the instability of $W_c$ for a given modulation configuration near the boundary, though the one-shot $W_c$ is always an integer in the thermodynamic limit. We take $v=1.1$ as an example shown in \textbf{Figure~\ref{Fig2}}d, the moderate slowly varying modulation $\Delta$ leads to a change of the nontrivial winding number in the region $\Delta \in (1,2)$, meanwhile the disorder-averaged zero modes are also detected. When $v>2$, the TAI phase disappears, no matter how strong the modulation amplitude is, as can be seen in \textbf{Figure~\ref{Fig2}}a. Our numerical results show that the slowly varying intracell modulation will evoke the TAI phase in a finite region for $v>1$.

According to Refs. \cite{Hughes2014,Meier2018,DZhang2020,Longhi2020}, the topological transition is accompanied by the divergence of the localization length of the zero modes. Hence, one can obtain the phase diagram by studying the localization length of the zero modes. For the zero modes, the Schr\"{o}dinger equation of the SSH model (\ref{eq1}) $H\psi=0$ reads:
\begin{eqnarray}
w\psi_{j,B}+v_{j+1}\psi_{j+1,B}&=&0 \notag \\
v_j\psi_{j,A}+w\psi_{j+1,A}&=&0,
\label{eq5}
\end{eqnarray}
where $\psi_{j,A}$($\psi_{j,B}$) is the probability amplitude of the zero mode on the sublattice site
$A(B)$ in the $j$-th lattice cell. By solving the coupled equations, one has $\psi_{n+1,A}=(-1)^n\prod_{j=1}^{n} (v_j/w) \psi_{1,A}$, leading to the localization length $\lambda$ of the zero modes given by \cite{Longhi2020}
\begin{eqnarray}
\lambda^{-1}&=&-\lim_{L\to\infty}\frac{1}{L}\ln\left|\frac{\psi_{L+1,A}}{\psi_{1,A}}\right| \notag \\
&=&\lim_{L\to\infty}\frac{1}{L}\left|\sum_{j=1}^{L}\ln{|v+\Delta\cos(2\pi \beta j^u)|}\right|.
\label{eq6}
\end{eqnarray}
According to Weyl's equidistribution theorem \cite{Weyl1916,Choe1993}, we can use the ensemble average to evaluate the last expression
\begin{equation}
\lambda^{-1}=\left|\frac{1}{2\pi}\int^{\pi}_{-\pi}dq\ln{|v+{\Delta}\cos q|} \right|.
\label{eq7}
\end{equation}
The integration can be performed straightforwardly as
\begin{equation}
\lambda^{-1}=\begin{cases}\ln{\frac{v+\sqrt{v^2-\Delta^2}}{2}} \quad & v>\Delta, \\
	\ln{\frac{\Delta}{2}} \quad & v<\Delta. \\
	\end{cases}
\label{eq8}
\end{equation}
The divergence of this localization length $\lambda$, i.e $\lambda^{-1}\to 0$, gives the two critical lines
\begin{align}
	v &= \Delta^2/4+1 \quad & v>\Delta, \notag \\
	\Delta &= 2 \quad & v<\Delta.
\label{eq9}
\end{align}
The localized critical points match the topological phase transition points shown in \textbf{Figure~\ref{Fig2}}a. Numerically the value of $\lambda^{-1}$ for the $L$-th eigenstate is computed by \cite{Wangg2020PRL}
\begin{eqnarray}
\lambda^{-1}&=&\lim_{L\to\infty}\frac{1}{L}||T||,
\end{eqnarray}
where $\|\cdot\|$ denotes the norm of the total transfer matrix $T=\prod_{j=2}^{L}T_jT_1$ with
\begin{eqnarray}
T_j&=&\left[\begin{array}{cc}
\frac{E_{L}^2-v_j^2-1}{v_j} & -\frac{v_{j-1}}{v_j} \\
1 & 0 \\
\end{array}
\right]
\end{eqnarray}
and
\begin{eqnarray}
T_1&=&\left[\begin{array}{cc}
\frac{(E_{L}^2-v_1^2)}{v_1}& -1\\
1 & 0\\
\end{array}
\right].
\end{eqnarray}
The numerical result of $\lambda^{-1}$ as a function of $v$ and $\Delta$ by means of the transfer matrix method \cite{Hughes2014,MacKinnon1983} is shown in \textbf{Figure~ \ref{Fig2}}b, and the two diverging critical lines match with the analytic solutions {Equation (\ref{eq9}) pretty well. Moreover, for $u=1$, the modulation is reduced to the AA-type, the topological phase diagram is the same (see Appendix A).

For $\Delta=0$, the system reduces to the original SSH model. When the intracell hopping strength $v$ exceeds the intercell hopping strength $w$, the system undergoes a topological phase transition from a topologically nontrivial phase into a trivial phase accompanied by a jump of the winding number from 1 to 0. However, when one introduces a slowly varying modulation for $u\ne 0$ and $\Delta \ne 0$, the topologically nontrivial phase, i.e. the TAI phase, emerges in the trivial regime of the original SSH model, which is shown in \textbf{Figure~ \ref{Fig2}}a for non-zero $\Delta$.

\section{\label{sec.4}Localization transition and mobility edges}
\begin{figure}[tbp]
 \includegraphics[width=0.5\textwidth]{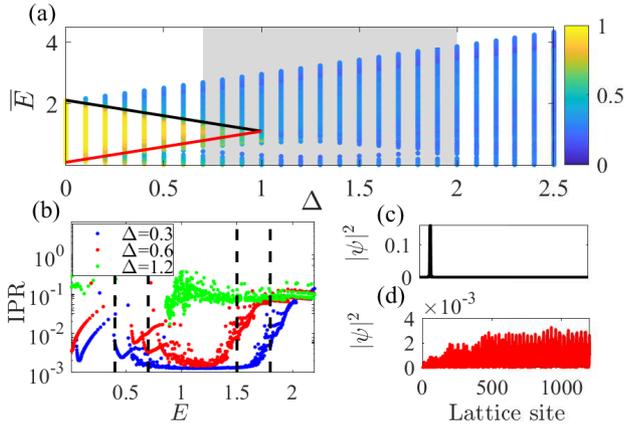}
  \caption{(Color online) (a) The disorder-averaged fractal dimension $\overline{\mathrm{\Gamma}}$ of different eigenstates as a function of the corresponding {$\overline{E}$} and the modulation strength $\Delta$ {with $N_c=100$ disorder realizations}. The colorbar indicates the magnitude of {$\overline{\mathrm{\Gamma}}$}. The black and red solid lines represent the mobility edges given in Equation (\ref{eq14}). The shaded regime represents the TAI regime. (b) The distribution of the ${\mathrm{IPR}}$ as a function of the ${E}$ with different $\Delta$ {with $\phi=0$}. The black dashed lines correspond to the mobility edges given in Equation (\ref{eq14}). The eigenstates spatial distributions with $\phi=0$ and $\Delta=0.5$ are shown in (c) for ${E}=1.9695$ satisfying $E>E_{c1}$ and (d) for ${E}=1.1108$ satisfying $E_{c2}<E<E_{c1}$, respectively, where $E_{c1}=1.6$ and $E_{c2}=0.6$ correspond to the mobility edges shown in Equation (\ref{eq14}) with $\Delta=0.5$.  Here, only ${E}> 0$ states are shown for $\beta=(\sqrt{5}-1)/2$, $v=1.1$ and $L=600$.}
\label{Fig3}
\end{figure}
\begin{figure}[tbp]
 \includegraphics[width=0.5\textwidth]{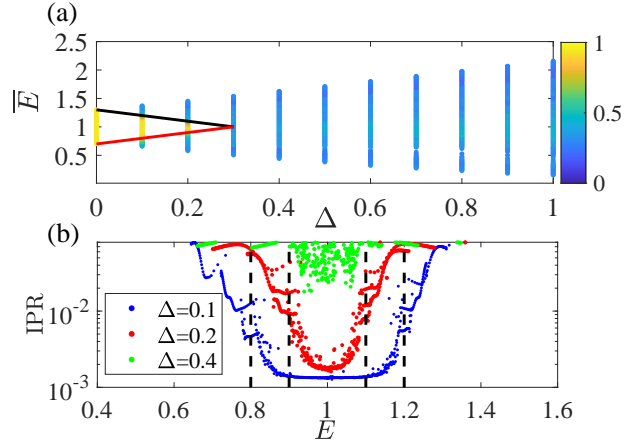}
  \caption{(Color online) (a) The disorder-averaged fractal dimension $\overline{\mathrm{\Gamma}}$ of different eigenstates as a function of the corresponding $\overline{E}$ and $\Delta$ with $N_c=100$ disorder realizations. The colorbar indicates the magnitude of $\overline{\mathrm{\Gamma}}$. The black and red solid lines represent the mobility edges given in Equation (\ref{eq17}). (b) The distribution of the IPR as a function of the eigenenergy $E$ with different $\Delta$ with $\phi=0$. The black dashed lines correspond to the positions of mobility edges given in Equation (\ref{eq17}). Here, only $E> 0$ states are considered, $\beta=(\sqrt{5}-1)/2$, $v=0.3$ and $L=600$.}
\label{Fig4}
\end{figure}

A common opinion is that when the system is in the TAI phase, all states are localized, showing Anderson's localization phenomenon. In this section, we study another aspect of the localized nature of the TAI phase, i.e. the access of mobility edges to the TAI phase in the SSH model under a slowly varying incommensurate modulation. We follow the method in Ref. \cite{He2021} to derive the expression of mobility edges for our slowly varying model.
Due to the chiral symmetry, the system's spectrum is symmetric about $E=0$. To simplify, we only focus on the upper band($E>0$), and the lower one displays identical behaviors.

The localization in the slowing varying model exhibits different features for two cases of the site-independent intracell hopping strength $v$. As following results imply, the delocalization-localization transition is not connected to a topological phase transition in both regimes.

\textbf{(a)} For $v>1$, the mobility edges are given by (see Appendix B for details):
\begin{equation}
E_{c} = v\pm\left(\Delta-1\right).
\label{eq14}
\end{equation}
Here and in Equation (\ref{eq17}) $+$($-$) denotes the upper(lower) mobility edge marked by the black(red) solid line shown in \textbf{Figure~\ref{Fig3}}a, respectively.
We find that $\Delta_c=1$ is the localization transition point. When $\Delta>\Delta_c$, all eigenstates become localized, and there exist mobility edges in the regime $0<\Delta<\Delta_c$. According to the topological phase diagram \textbf{Figure~\ref{Fig2}}a, we find that the TAI phase is compatible with the regime hosting mobility edges.
To characterize the mobility edges, we define the fractal dimension of the wave function as following:
\begin{equation}
	{\mathrm{\Gamma}} = -\lim_{L\to\infty}\frac{{\ln{{\mathrm{IPR}}}}}{\ln{(2L)}},
	\label{eq18}
\end{equation}
where the inverse participation ratio ${\mathrm{IPR}}=\sum_{j=1}^{L}\sum_{\alpha}|\psi_{j,\alpha}^{n}|^4$. Further we can define the disorder-averaged fractal dimension $\overline{\mathrm{\Gamma}}=1/N_c\sum_{c=1}^{N_c}\mathrm{\Gamma}^c$.
It is known that $\Gamma \to 1$ for extended states and $\Gamma \to 0$ for localized ones \cite{Wangg2020PRL}. In \textbf{Figure~\ref{Fig3}}a, we show the disorder-averaged fractal dimension $\overline{\Gamma}$ of the eigenstates as a function of the modulation strength $\Delta$ for $v=1.1$. The solid lines represent the mobility edges defined by Equation (\ref{eq14}), and the shaded regime denotes the region with TAI. One can see that there exists an overlap between the TAI regime and the extended states regime with mobility edges. To clarify the mobility edges more clearly, \textbf{Figure~\ref{Fig3}}b plots the ${\mathrm{IPR}}$ as a function of {the eigenenergy ${E}$} for different $\Delta$ with $v=1.1$. The magnitude of the ${\mathrm{IPR}}$ jumps from the order of magnitude of $10^{-2}$(a typical value for the localized states \cite{Tong2018PRB}) to $10^{-3}$(a typical value scales as $1/L$ for the extended states \cite{Tong2018PRB}), when {the eigenenergy ${E}$} crosses some critical values for $\Delta=0.3$ and $0.6$. This clearly signals the transition between localized and extended states, and these critical values correspond to the positions of the mobility edges in the spectrum marked by the black dotted lines in \textbf{Figure~\ref{Fig3}}b defined by Equation (\ref{eq14}). When $\Delta=1.2>\Delta_c$, the ${\mathrm{IPR}}$ magnitude for all eigenstates is up to  $10^{-2}$, which indicates there is no mobility edges in the spectrum. Furthermore, the mobility edges can also be confirmed intuitively by the spatial distributions of the wave functions in \textbf{Figure~\ref{Fig3}}c and \textbf{Figure~\ref{Fig3}}d. The wave functions are localized or extended when their eigenvalues satisfy ${E}>v-\left(\Delta-1\right)$ and $v+\left(\Delta-1\right)<{E}<v-\left(\Delta-1\right)$, respectively.

\textbf{(b)} For $0<v<1$, the mobility edges are instead
\begin{equation}
E_{c}= 1\pm\left(v-\Delta\right).
\label{eq17}
\end{equation}
\textbf{Figure~\ref{Fig4}}a shows the disorder-averaged fractal dimension $\overline{{\Gamma}}$ of the eigenstates as a function of disorder-averaged eigenenergy $\overline{{E}}$ and modulation strengthen $\Delta$ for $v=0.3$. In the regime $0<\Delta<v$, as expected from the analytical results, $\overline{\Gamma}$ approximately exhibits sharp jumps from $1$ to $0$ for energies upper or lower lines defined by Equation (\ref{eq17}). When $\Delta > v$, the mobility edges vanish, and all the states are localized.  \textbf{Figure~\ref{Fig4}}b plots the IPR as a function of the eigenenergy for different $\Delta$ with $v=0.3$, where the jumping points denote the positions of the mobility edges.

\section{\label{sec.5}Conclusion}
In summary, we study the topological phase transition and localization transition of a SSH model with a slowly incommensurate modulation. We numerically and analytically obtain the topological phase diagram, and our results imply that a slowly vary incommensurate modulation can induce the TAI. Different from the random disorder induced the TAI, the mobility edges can enter into the TAI regime in our case.

Note added: on submission of this manuscript, we notice a very recent preprint arXiv:2201:00988 entitled 'Topological Anderson insulators with different bulk states in quasiperiodic chains' \cite{Tangg2022}. The authors studied a similar problem, and main results of the TAI regime with the intermediate phase were obtained. In this paper, we derive the exact expression of mobility edges in this TAI phase.

\section*{Acknowledgements}
This work was supported by the National Natural Science Foundation of China (12074340), the NSFC (Grants No. 11604188 and No. 12147215), Fundamental Research Program of Shanxi Province (Grant No. 20210302123442), Beijing National Laboratory for Condensed Matter Physics, and STIP of Higher Education Institutions in Shanxi under Grant No. 2019L0097. This work is also supported by NSF for Shanxi Province Grant No. 1331KSC.

\section*{Conflict of Interest}
The authors declare no conflict of interest.

\appendix
\begin{center}
\section*{ Appendix A}
\end{center}
\renewcommand{\theequation}{A\arabic{equation}}
\setcounter{equation}{0}
\renewcommand\thefigure{A\arabic{figure}}
\setcounter{figure}{0}
\section*{AA-type disorder}

\begin{figure}[h]
 \includegraphics[width=0.45\textwidth]{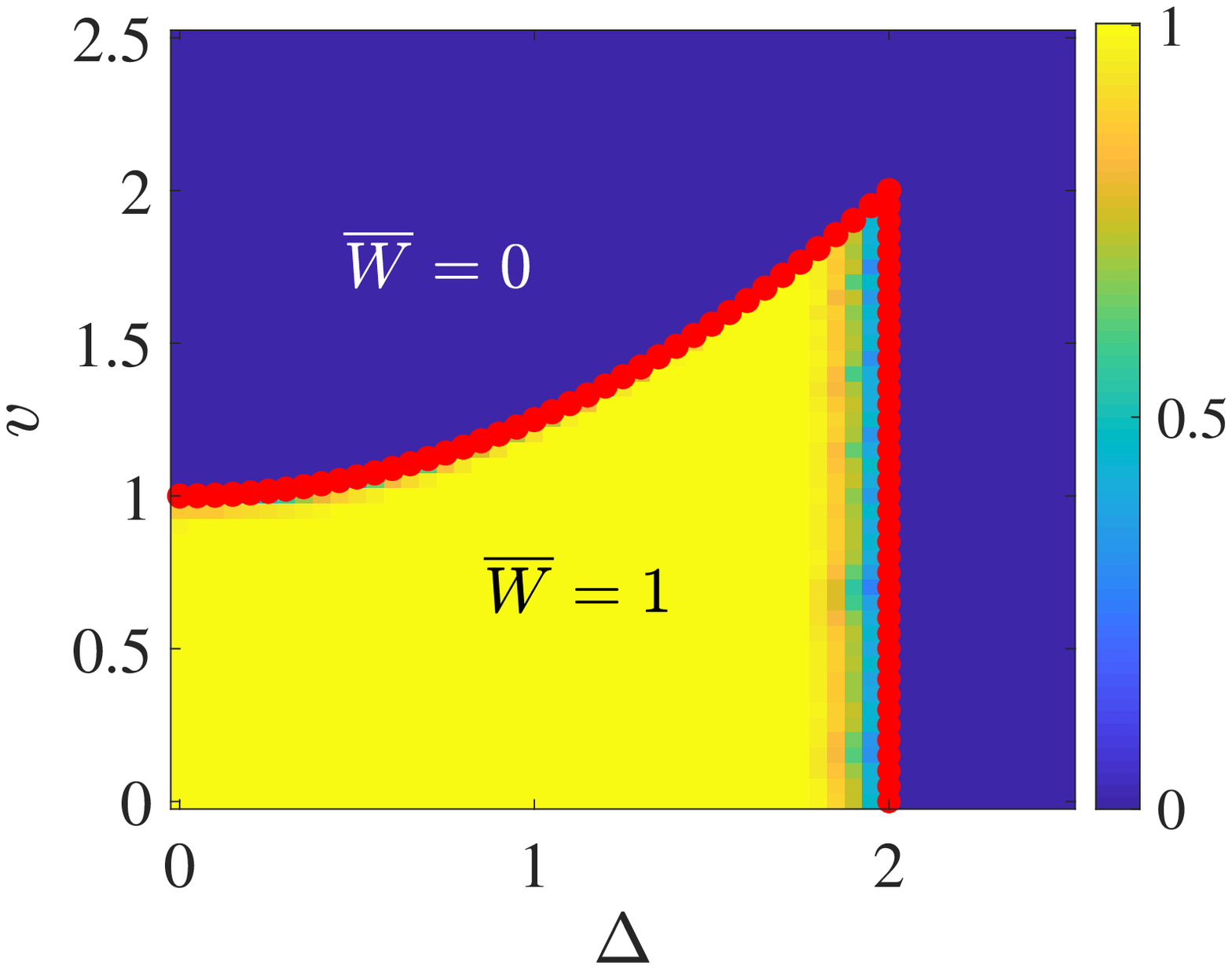}
  \caption{(Color online) Disorder-averaged winding number $\overline{W}$ as a function of the modulation strength $\Delta$ and the amplitude of the intracell hopping $v$ with $L=300$ and 100 disorder  realization. Two red dot lines represent the analytic critical lines Equation (\ref{eq9}). Other parameters: $\beta=(\sqrt{5}-1)/2$ and $u=1$.}
\label{Fig7}
\end{figure}

When $u=1$, $\Delta_j$ is reduced to the AA-type modulation. \textbf{Figure \ref{Fig7}} shows the topological phase diagram characterized by $\overline{W}$ versus the disorder strength $\Delta$ and intracell hopping $v$, which is the same with the result in \textbf{Figure~\ref{Fig2}}a. As shown in \textbf{Figure~\ref{Fig7}}, the TAI phase also can be introduced by AA-type modulation and topological critical points still can be described by Equation (\ref{eq9}), which labeled by two red dot lines, shown in \textbf{Figure~\ref{Fig7}}.

\begin{figure}[h]
 \includegraphics[width=0.45\textwidth]{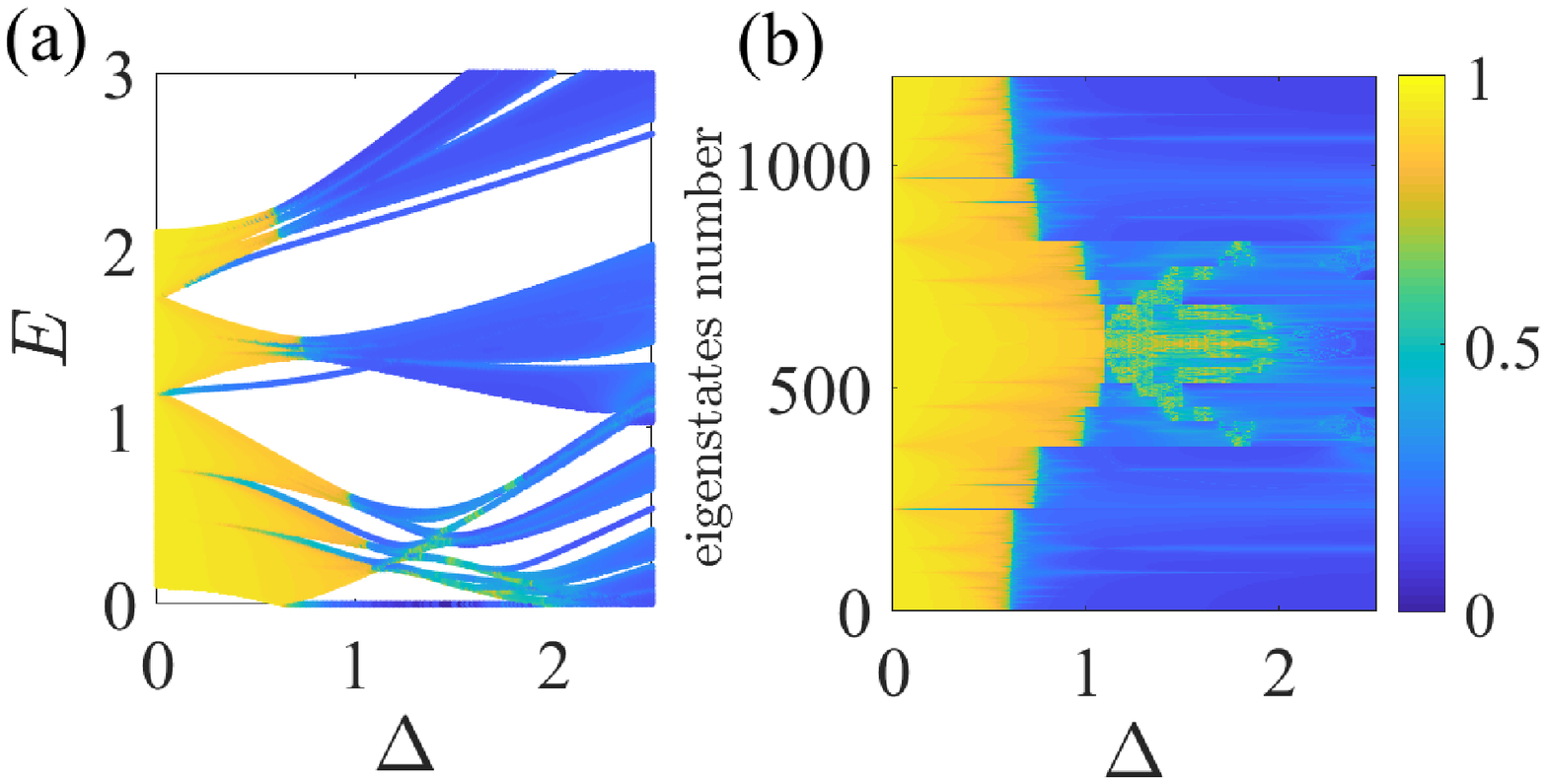}
  \caption{(Color online)(a) Fractal dimension $\Gamma$ of different eigenstates as a function of the corresponding eigenvalues and  $\Delta$. (b) The $\Gamma$ associated to the
  eigenstate number indices as a function $\Delta$. Other parameters: $\beta=(\sqrt{5}-1)/2$, $u=1$, $\phi=0$, $v=1.1$ and $L=600$.}
\label{Fig8}
\end{figure}

To figure out the localization property in the TAI phase, we display {the Fractal dimension ${\Gamma}$} of different eigenstates as a function of the corresponding ${E}$ and  $\Delta$ for $v=1.1$ in \textbf{Figure~\ref{Fig8}}a. One can see that some eigenstates are localized and the rest are extended, which indicates the mobility edges exist in system. By increasing $\Delta$, we can find an intermediate region consisting of both extended, localized, and even critical eigenstates exists in spectrum. For a clear comparison, the ${\Gamma}$ associated to eigenstates as a function of $\Delta$ is shown in \textbf{Figure~\ref{Fig8}}b. With the increase of $\Delta$, more extended states become localized, shown in \textbf{Figure~\ref{Fig8}}a and b. And an intermediate region with a mixture of extended, localized, and critical eigenstates emerge in $1.1<\Delta<2$. When the AA-type modulation is introduced in the system, the localization feature becomes very complex. So we can't get the analytic expression of mobility edges.

\section*{B Derivation of the analytical expression for mobility edges}
\renewcommand{\thesubsection}{B}
\renewcommand{\theequation}{B\arabic{equation}}
\renewcommand{\thetable}{B\Roman{table}}
\setcounter{equation}{0}
\renewcommand\thefigure{B\arabic{figure}}
\setcounter{figure}{0}

\begin{table}[h]
\centering{
\begin{tabular}{|c|c|c|c|c|}
\hline
\multicolumn{1}{|l|}{} & $v$                & $\Delta$                &$\left(v,\Delta\right)$        & the smallest energy overlaps
 \\ \hline
{\rm{\uppercase\expandafter{\romannumeral1}}}     & {$v$ \textless{}1}        & {$\Delta$\textless{}1}        & $\Delta$\textless{}$v$    & $\left(|{E}_3|,|{E}_2|\right)$                                                    \\ \cline{4-5}
                       &                                       &                                       & $\Delta$\textgreater{}$v$           &$\left(|{E}_2|,|{E}_3|\right)$                                                    \\ \hline
{\rm{\uppercase\expandafter{\romannumeral2}}}     & {$v$\textgreater{}1}     & \multicolumn{1}{l|}{$\Delta$\textless{}1}    & {}          &$\left(|{E}_1|,|{E}_2|\right)$                                  \\ \cline{3-3} \cline{5-5}
                       &                                       & \multicolumn{1}{l|}{$\Delta$\textgreater{}1} &                              & $\left(|{E}_2|,|{E}_1|
                       \right)$                                                   \\ \hline
{\rm{\uppercase\expandafter{\romannumeral3}}}     & {$v$\textless{}1}        & {$\Delta$\textless{}1}        & $\Delta$\textgreater{}$v$   &$\left(|{E}_2|,|{E}_3|\right)$                                                     \\ \cline{4-5}
                       &                                       &                                       & $\Delta$\textless{}$v$                & $\left(|{E}_3|,|{E}_2|\right)$                                                  \\ \cline{2-5}
                       & \multicolumn{1}{l|}{$v$\textless{}1}    & {$\Delta$\textgreater{}1}     & {$\Delta$\textgreater{}$v$} &$\left(|{E}_1|,|{E}_3|\right)$                                                    \\ \cline{2-2} \cline{5-5}
                       & \multicolumn{1}{l|}{$v$\textgreater{}1} &                                       &                                   & $\left(|{E}_3|,|{E}_1|\right)$                                                      \\ \cline{2-5}
                       & {$v$\textgreater{}1}     & \multicolumn{1}{l|}{$\Delta$\textless{}1}    & {$\Delta$\textless{}$v$}    & $\left(|{E}_1|,|{E}_2|\right)$                                                    \\ \cline{3-3} \cline{5-5}
                       &                                       & \multicolumn{1}{l|}{$\Delta$\textgreater{}1} &                                   & $\left(|{E}_2|,|{E}_1|\right)$                                                      \\ \hline
{\rm{\uppercase\expandafter{\romannumeral4}}}     & \multicolumn{1}{l|}{$v$\textless{}1}    & {$\Delta$\textgreater{}1}     & {}                 & $\left(|{E}_1|,|{E}_3|\right)$                                                    \\ \cline{2-2} \cline{5-5}
                       & \multicolumn{1}{l|}{$v$\textgreater{}1} &                                       &                                   & $\left(|{E}_3|,|{E}_1|\right)$                                                     \\ \hline
\end{tabular}}
\caption{The smallest energy overlaps of extended states in ${\rm{\uppercase\expandafter{\romannumeral1}}}$, ${\rm{\uppercase\expandafter{\romannumeral2}}}$, ${\rm{\uppercase\expandafter{\romannumeral3}}}$ and ${\rm{\uppercase\expandafter{\romannumeral4}}}$ regions.  }
\label{Ta2}
\end{table}

\begin{figure}[h]
	\includegraphics[width=0.5\textwidth]{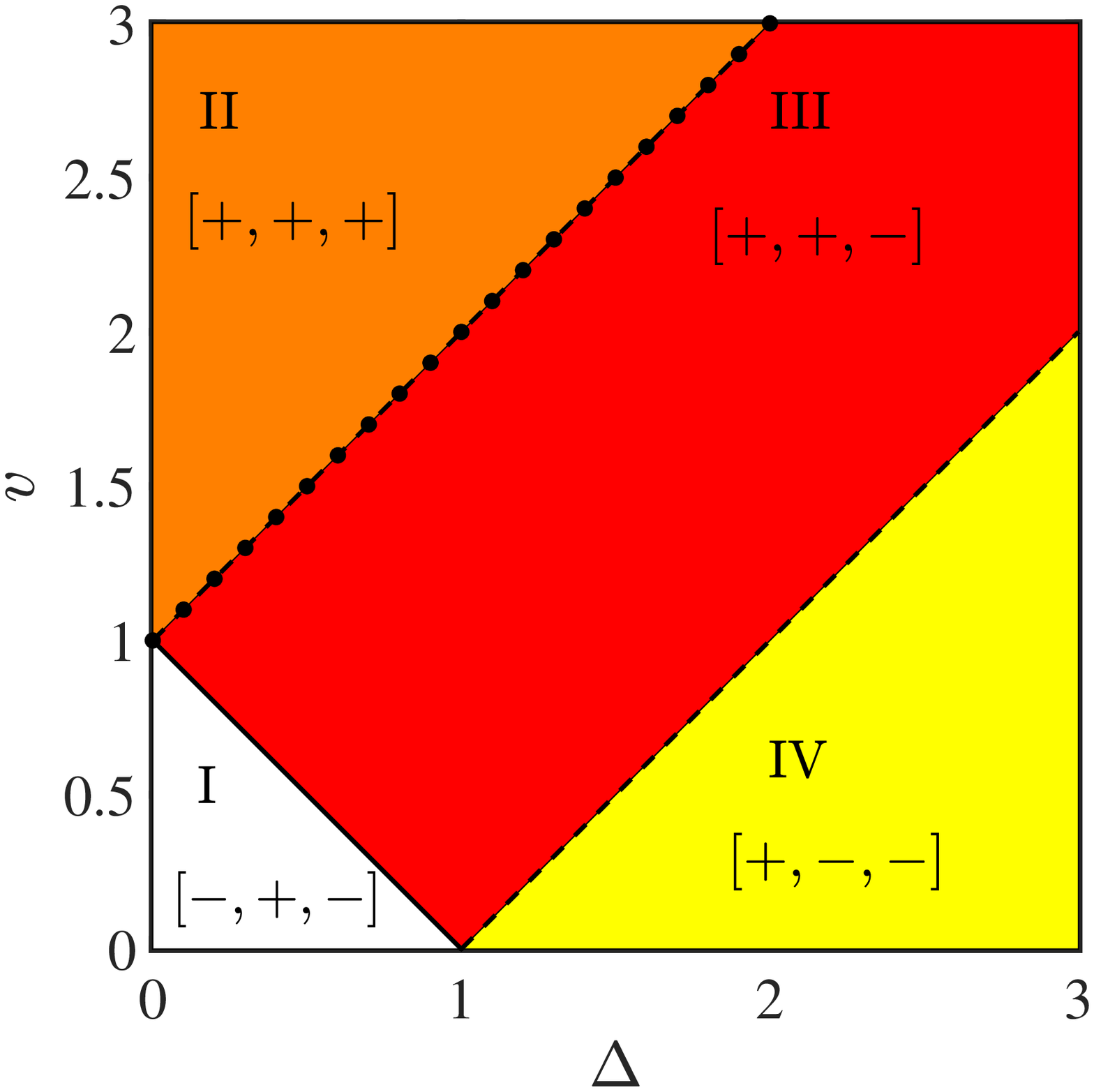}
	\caption{(Color online){The sign of ${E}_1$, ${E}_2$ and ${E}_3$ in the $v-\Delta$ plane.}}
     \label{Fig5}
\end{figure}

In this Appendix, we present a detailed derivation of the expression of mobility edges Equations (\ref{eq14}) and (\ref{eq17}) in details following the method proposed in Ref. \cite{He2021}. To do this, we approximate the slowly varying modulation model with an ensemble of different periodic models $M_a$. 
And the extended state regime in the spectrum of the slowly varying modulation model corresponds to the range of energy bands of these periodic models $M_a$. Thus, the mobility edges can be determined as the smallest overlaps of these ranges.

We first assume the slowly varying incommensurate modulation $\Delta_j$ may take a series of constants $\Delta_a(\Delta_a\in[-\Delta, \Delta])$ in the thermodynamic limit, and the system becomes an ensemble of periodic models $M_a$. Under periodic boundary conditions(PBCs) with $c^\dagger_{L+1,\alpha}=c^\dagger_{1,\alpha}$, the energy spectra of $M_a$ can be obtained by the tight-binding approximation as follows:
\begin{eqnarray}
E&=& \pm\sqrt{(v+\Delta_a)^2+1+2(v+\Delta_a)\cos k},
\end{eqnarray}
where $k\in(0,2\pi]$ is the wave number. Here we only focus on the upper band($E>0$) for positive $v$ and $\Delta$.

The extended state regime for $M_a$ is just the range of the energy band:
\begin{eqnarray}
|v+\Delta_a-1|<E<|v+\Delta_a+1|
\end{eqnarray}
or
\begin{eqnarray}
|v+\Delta_a+1|<E<|v+\Delta_a-1|.
\end{eqnarray}
There exists two limit situations for $\Delta_a$. When $\Delta_a=\Delta$, we have
\begin{eqnarray}
|{E}_1|<E<|{E}_4|,
\label{aaaa}
\end{eqnarray}
with $E_1=v+\Delta-1$ and $E_4=v+\Delta+1$.
The other limit is $\Delta_a=-\Delta$, for that we have
\begin{eqnarray}
|E_2|<E<|E_3|
\end{eqnarray}
or
\begin{eqnarray}
|E_3|<E<|E_2|,
\end{eqnarray}
with $E_2=v-\Delta+1$ and $E_3=v-\Delta-1$.
To determine the smallest overlaps of energy bands of $M_a$, one shall first sort the values of $|E_1|$, $|E_2|$, and $|E_3|$. We first determine the sign of $E_1$, $E_2$, and $E_3$ in the $v$-$\Delta$ plane. As shown in \textbf{Figure~\ref{Fig5}}, the $v$-$\Delta$ plane is divided into four regions by three lines corresponding to $E_1=0$, $E_2=0$, and $E_3=0$, respectively, each labeled with sign combinations of $E_1$, $E_2$, and $E_3$. In region  ${\rm{\uppercase\expandafter{\romannumeral1}}}$, for instance, we have $|{E}_1|<|E_3|$ and $
|E_1|<|E_2|$. When $\Delta<v$, $|E_1|<|E_3|<|E_2|<|E_4|$, and the smallest energy overlaps for the extended states is $E \in (|{E}_3|,|{E}_2|)$. Thus, the mobility edges reads
\begin{eqnarray}
E_c=1\pm(v-\Delta).
\end{eqnarray}
When $\Delta>v$, $|{E}_1|<|{E}_2|<|{E}_3|<|E_4|$. The smallest energy overlaps of the extended states is given by
$E \in (|{E}_2|,|{E}_3|)$, and the expression of mobility edges is
\begin{eqnarray}
E_c=1\pm(v-\Delta).
\end{eqnarray}
For the regions ${\rm{\uppercase\expandafter{\romannumeral2}}}$,  ${\rm{\uppercase\expandafter{\romannumeral3}}}$ and ${\rm{\uppercase\expandafter{\romannumeral4}}}$,
the smallest energy overlaps can be obtained by the same procedure. Table \ref{Ta2} lists the smallest energy overlaps for different $v$ and $\Delta$ in four regions.

According to Table \ref{Ta2}, there are three forms for mobility edges, {\it i.e.}, $E_{c1}=|{E}_1|$, $E_{c2}=|{E}_2|$, and $E_{c3}=|{E}_3|$.
\begin{figure}[h]
 \includegraphics[width=0.5\textwidth]{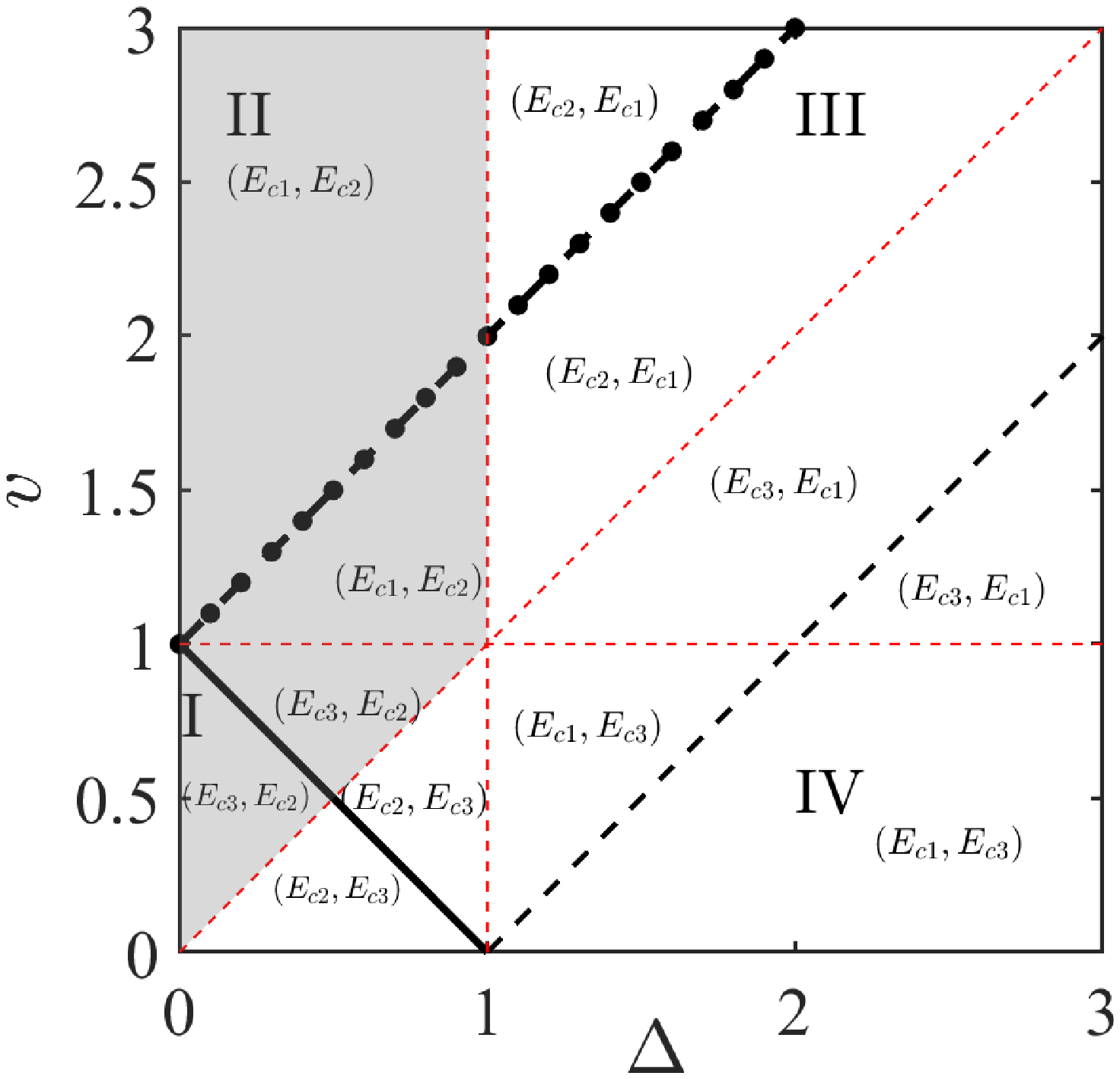}
\caption{(Color online){The distribution of the smallest energy overlaps for different $v$ and $\Delta$ in different regions. The shaded area indicates the regime of the system with mobility edges.}}
\label{Fig6}
\end{figure}
In \textbf{Figure~\ref{Fig6}}, we show the region of the smallest energy overlaps of the extended states in the $v$-$\Delta$ plane for $M_a$. We find that the smallest energy overlaps of the extended states change from $(E_{c3},E_{c2})$ to $(E_{c2},E_{c3})$, when $\Delta$ goes beyond $v$ in the region $\rm{\uppercase\expandafter{\romannumeral1}}$. At $E_{c2}=E_{c3}$, the extended state regime shrinks to a point with $\Delta=v$. In the regions $\rm{\uppercase\expandafter{\romannumeral2}}$ and $\rm{\uppercase\expandafter{\romannumeral3}}$, the localization transition occurs at $\Delta=1$ and $\Delta=v$. The shaded area in \textbf{Figure~\ref{Fig6}} indicates the regime of the system where the mobility edges actually exist. Outside this region, all states are kept localized. As a conclusion, when $v>1$, $E_{c} = v\pm\left(\Delta-1\right)$, and the localization transition occurs at $\Delta=1$.
And for $0<v<1$, $E_{c}= 1\pm\left(v-\Delta\right)$, and the localization transition occurs at $\Delta=v$.




\end{document}